\documentstyle[a4,12pt]{article}
\renewcommand{\baselinestretch}{1.2}
\def\thefootnote{\fnsymbol{footnote}}
\renewcommand{\theequation}{\thesection.\arabic{equation}}

\begin{document}
\parskip=5pt plus 1pt minus 1pt

\begin{flushright}
{\bf DPNU-97-36}\\
{July, 1997}
\end{flushright}

\vspace{0.2cm}

\begin{center}
{\Large\bf Search for New Physics in $CP$-violating $B$ Decays}
\end{center}

\vspace{0.3cm}

\begin{center}
{\bf A.I. Sanda} \footnote{Electronic address:
sanda@eken.phys.nagoya-u.ac.jp}
~ and ~ 
{\bf Zhi-zhong Xing} \footnote{Electronic address: xing@eken.phys.nagoya-u.ac.jp}
\end{center}

\begin{center}
{\it Department of Physics, Nagoya University, Chikusa-ku, Nagoya 464-01, Japan}
\end{center}

\vspace{3cm}

\begin{abstract}

We consider three possible scenarios of new physics in $B^0_d$-$\bar{B}^0_d$ mixing
and propose a simple framework for analyzing their effects.
This framework allows us to study the $CP$ asymmetry in
semileptonic $B_d$ decays (${\cal A}_{\rm SL}$) and those in
nonleptonic transitions such as 
$B_d \rightarrow J/\psi K_S$ and $B_d \rightarrow \pi^+\pi^-$.
Numerically we find that new physics may enhance the magnitude of 
${\cal A}_{\rm SL}$ up to the percent level within the appropriate 
parameter space. So measurements of ${\cal A}_{\rm SL}$ and its
correlation with other $CP$ asymmetries will serve as a sensitive 
probe for new physics in $B^0_d$-$\bar{B}^0_d$ mixing.  
\end{abstract}

\vspace{2cm}

\begin{center}
PACS number(s): 11.30.Er, 12.15.Ff, 13.25.+m
\end{center}

\newpage

\section{Introduction}
\setcounter{equation}{0}

The $B$-meson factories under construction at KEK and SLAC will
provide a unique opportunity to study $CP$ violation in weak
$B$ decays. One type of $CP$-violating signals, arising from
$B^0_d$-$\bar{B}^0_d$ mixing, is expected to manifest itself in
the decay rate asymmetry between two semileptonic channels
$B^0_d\rightarrow l^+\nu^{~}_l X^-$ and $\bar{B}^0_d\rightarrow
l^-\bar{\nu}^{~}_l X^+$. This $CP$ asymmetry, denoted as ${\cal
A}_{\rm SL}$ below, has been estimated to be of the order $10^{-3}$
within the standard model (SM) \cite{Wolfenstein88,Lusignoli89}.
Obviously the magnitude of ${\cal A}^{\rm SM}_{\rm SL}$ is too small to be
measured in the first-round experiments of any $B$ factory. The
current experimental constraint on ${\cal A}_{\rm SL}$ is very
rough: $|{\cal A}_{\rm SL}| < 0.18$ at the $90\%$
confidence level \cite{CLEO93/CDF97/OPAL97,PDG96}. Nevertheless, one expects that this
limit will be greatly improved once $B$-meson factories start collecting data.

Importance of the $CP$ asymmetry ${\cal A}_{\rm SL}$ has been
repeatedly emphasized
for the purpose of searching for new physics (NP) in the $B$-meson
system (see, e.g., Refs. \cite{Sanda87,Maiani92}).
The presence of NP in $B^0_d$-$\bar{B}^0_d$ mixing
might enhance the magnitude of ${\cal A}_{\rm SL}$ to an
observable level. For instance, $|{\cal A}_{\rm SL}| \sim 10^{-2}$ 
can be achieved from a specific superweak model proposed in
Ref. \cite{Maiani92}, where the Kobayashi-Maskawa (KM)
matrix is assumed to be real and the $CP$-violating phase
comes solely from the NP. It is therefore worthwhile to search for 
$CP$ violation in semileptonic $B_d$ decays
at $B$-meson factories, in order to determine or
constrain both the magnitude and 
the phase information of possible NP in $B^0_d$-$\bar{B}^0_d$ mixing.
The $CP$-violating phase of NP may also be isolated from measuring
$CP$ asymmetries in some $B_d$ decays into hadronic $CP$ eigenstates,
such as $B_d \rightarrow J/\psi K_S$ and $B_d \rightarrow \pi^+\pi^-$
modes \cite{AngleNP97,Xing96}.

In this paper we investigate the 
$CP$ asymmetry ${\cal A}_{\rm SL}$ and its correlation with the $CP$
asymmetries in $B_d \rightarrow J/\psi K_S$ and $B_d\rightarrow
\pi^+\pi^-$, based on three NP
scenarios for $B^0_d$-$\bar{B}^0_d$ mixing. Scenario (A)
allows $B^0_d$-$\bar{B}^0_d$ mixing to contain a real SM-like term and 
a complex superweak contribution \cite{Footnote1};
scenario (B) requires $B^0_d$-$\bar{B}^0_d$ mixing to include the normal
SM effect with an additional real superweak contribution 
\cite{Wolfenstein93}; and scenario (C) is a general
case in which both the SM and NP contributions to
$B^0_d$-$\bar{B}^0_d$ mixing are complex ($CP$-violating).

For each scenario, we first propose a simple parametrization of the NP effect,
and then calculate $CP$-violating asymmetries in the above mentioned $B$ decays.
Some numerical estimates for these $CP$ asymmetries in scenarios (A)
and (B) are also made. We find that
for all three scenarios the magnitude of ${\cal A}_{\rm SL}$ can
reach the percent level within the suitable parameter space.
Thus an experimental study of the correlation between $CP$ asymmetries in the
semileptonic and nonleptonic $B_d$ decays should impose useful
constraints on possible NP in $B^0_d$-$\bar{B}^0_d$ mixing.

The remainder of this paper is organized as follows. In section 2,
some necessary preliminaries for $B^0_d$-$\bar{B}^0_d$ mixing, $CP$
violation and the KM
unitarity triangle are briefly reviewed. The SM prediction for $CP$
violation in semileptonic $B_d$ decays, i.e., ${\cal A}^{\rm SM}_{\rm SL}$, is updated 
in section 3. Section 4 is devoted 
to parametrizing NP effects in $B^0_d$-$\bar{B}^0_d$ mixing and  
calculating ${\cal A}_{\rm SL}$ as well as its correlation with the $CP$
asymmetries in $B_d \rightarrow J/\psi K_S$ and $B_d \rightarrow
\pi^+\pi^-$, on the basis of three different NP scenarios. We 
numerically illustrate the allowed parameter space and $CP$
asymmetries for scenarios (A) and (B) in section 5. 
Finally some concluding remarks are given in section 6.

\section{Preliminaries}
\setcounter{equation}{0}

The mass eigenstates of $B^0_d$ and $\bar{B}^0_d$ mesons can be written,
in the assumption of $CPT$ invariance, as 
\begin{eqnarray}
|B_{\rm L}\rangle & = & p |B^0_d\rangle ~ + ~ q |\bar{B}^0_d\rangle \; ,
\nonumber \\
|B_{\rm H}\rangle & = & p |B^0_d\rangle ~ - ~ q |\bar{B}^0_d\rangle \; ,
\end{eqnarray}
where $p$ and $q$ are complex mixing parameters. In terms of the 
off-diagonal elements of the
$2\times 2$ $B^0_d$-$\bar{B}^0_d$ mixing Hamiltonian ${\bf M} -{\rm i}
{\bf \Gamma}/2$, we express the ratio $q/p$ as 
\begin{equation}
\frac{q}{p} \; =\; \sqrt{\frac{M^*_{12} - {\rm i}
\Gamma^*_{12}/2}{M_{12} - {\rm i} \Gamma_{12}/2}} \; .
\end{equation}
To a good approximation (i.e., $|M_{12}| \gg |\Gamma_{12}|$), 
the mass difference of $B_{\rm H}$ and
$B_{\rm L}$ (denoted by $\Delta M$) is related to $|M_{12}|$ 
through $\Delta M = 2 |M_{12}|$, and $q/p = \sqrt{M^*_{12}/M_{12}}$
holds.

The $CP$-violating asymmetry ${\cal A}_{\rm SL}$, for either 
incoherent or coherent decays of
$B^0_d$ and $\bar{B}^0_d$ mesons, is given by \cite{Hagelin81}
\begin{equation}
{\cal A}_{\rm SL} \; =\; \frac{|p|^4 ~ - ~ |q|^4}{|p|^4 ~ + ~ |q|^4}
\; =\; {\rm Im} \left ( \frac{\Gamma_{12}}{M_{12}}
\right ) \; ,
\end{equation}
At a $B$-meson factory, this signal can be extracted from 
the same-sign dilepton asymmetry on the $\Upsilon (4S)$ resonance.

Another observable, which is of particular interest for 
testing the KM mechanism of $CP$ violation, is the 
$CP$ asymmetry in $B^0_d$ vs $\bar{B}^0_d \rightarrow J/\psi K_S$
modes \cite{Sanda80}:
\begin{equation}
{\cal A}_{\psi K} \; =\; - ~ {\rm Im} \left ( \frac{q}{p} ~ 
\frac{V_{cb}V^*_{cs}}{V^*_{cb}V_{cs}} ~ \frac{q^*_K}{p^*_K} \right )
\; ,
\end{equation}
where the minus sign comes from the $CP$-odd eigenstate $J/\psi K_S$, and
$q^{~}_K/p^{~}_K$ describes the $K^0$-$\bar{K}^0$ mixing phase in 
the final state. Neglecting the tiny $CP$-violating effect measured from
$K^0$-$\bar{K}^0$ mixing, one can find that $q^{~}_K/p^{~}_K$ is
essentially unity in an appropriate phase convention, no matter
whether NP exists or not.  
Thus the above asymmetry turns out approximately to be
${\cal A}_{\psi K} = - {\rm Im} (q/p)$, if one adopts the Wolfenstein phase
convention \cite{Wolfenstein83} for the KM matrix.

In the neglect of penguin effects, measuring the $CP$ asymmetry in
$B^0_d$ vs $\bar{B}^0_d\rightarrow \pi^+\pi^-$ modes is also promising 
to probe the $CP$-violating weak phase \cite{Sanda80}:
\begin{equation}
{\cal A}_{\pi\pi} \; =\; {\rm Im} \left ( \frac{q}{p} ~
\frac{V_{ub}V^*_{ud}}{V^*_{ub} V_{ud}} \right ) \; .
\end{equation}
The correlation between ${\cal A}_{\pi\pi}$ and ${\cal A}_{\psi K}$ is 
sensitive to a variety of NP scenarios, such as superweak models (see,
e.g., Refs. \cite{Wolfenstein93,Gerard91}).

Within the SM, ${\cal A}_{\rm SL}$, ${\cal A}_{\psi K}$ and
${\cal A}_{\pi\pi}$ are related to the 
inner angles of the KM unitarity triangle 
formed by three vectors $\xi_i \equiv V^*_{ib}V_{id}$ ($i=u,c,t$) in
the complex plane (see Fig. 1 for illustration).
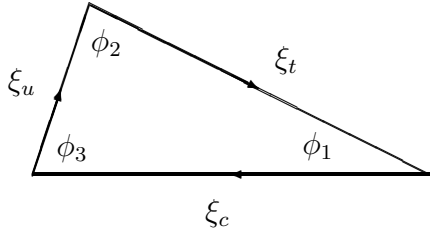
\begin{figure}[t]
\begin{picture}(400,160)(-90,210)
\put(80,300){\line(1,0){150}}
\put(80,300.5){\line(1,0){150}}
\put(230,300){\vector(-1,0){75}}
\put(230,300.5){\vector(-1,0){75}}
\put(150,285.5){\makebox(0,0){$\xi_c$}}
\put(80,300){\line(1,3){21.5}}
\put(80,300.5){\line(1,3){21.5}}
\put(80,299.5){\line(1,3){21.5}}
\put(80,300){\vector(1,3){10.75}}
\put(80,300.5){\vector(1,3){10.75}}
\put(80,299.5){\vector(1,3){10.75}}
\put(76,335){\makebox(0,0){$\xi_u$}}
\put(230,300){\line(-2,1){128}}
\put(230,300.5){\line(-2,1){128}}
\put(101.5,364.5){\vector(2,-1){64}}
\put(101.5,365){\vector(2,-1){64}}
\put(176,343.5){\makebox(0,0){$\xi_t$}}

\put(95,310){\makebox(0,0){$\phi_3$}}
\put(188,311){\makebox(0,0){$\phi_1$}}
\put(108,350){\makebox(0,0){$\phi_2$}}
\end{picture}
\vspace{-2.5cm}
\caption{\small Unitarity triangle ($\xi_u + \xi_c + \xi_t =0$) 
in the complex plane.}
\end{figure}
The sides $|\xi_u|$ and $|\xi_c|$ have been model-independently
measured, while the existence of NP may in general affect determination of the side 
$|\xi_t|$ from the rate of $B^0_d$-$\bar{B}^0_d$ mixing. 
In terms of the Wolfenstein parameters, one has
\begin{eqnarray}
\xi_u & \approx & A \lambda^3 \left ( \rho + {\rm i} \eta \right ) \;
, \nonumber \\
\xi_c & \approx & - A \lambda^3 \; , \nonumber \\
\xi_t & \approx & A \lambda^3 \left ( 1 - \rho - {\rm i} \eta \right ) 
\; .
\end{eqnarray}
For convenience in subsequent discussions, we define 
\begin{equation}
\chi \; \equiv \; \left | \frac{\xi_u}{\xi_c} \right | \; \approx \;
\sqrt{\rho^2 + \eta^2} \; ,
\end{equation}
which is a NP-independent quantity.

Note that in this work we only assume the kind of NP which does not
involve extra quark(s). This requirement implies that quark mixing
remains to be described by the $3\times 3$ KM matrix even though NP is 
present in $B^0_d$-$\bar{B}^0_d$ mixing. For some interesting extensions of the
SM, e.g., the supersymmetric models, the above assumption is of course satisfied.
We further assume that the penguin and tree-level contributions
to $B_d \rightarrow \pi^+\pi^-$ may be separated from each other 
using some well-known techniques \cite{Sanda97,Gronau96/Deshpande97},
thus ${\cal A}_{\pi\pi}$ is useful for probing the $CP$-violating 
weak phase(s).

\section{$CP$ asymmetries in the SM}
\setcounter{equation}{0}

Within the SM, both $M_{12}$ and $\Gamma_{12}$ can be reliably
calculated in the box-diagram approximation. Due to the dominance of
the top-quark contribution, $M^{\rm SM}_{12}$ reads \cite{Cheng82,Buras84}
\begin{equation}
M^{\rm SM}_{12} \; =\; \frac{G^2_{\rm F} ~ B_B ~ f^2_B ~ M_B ~ m^2_t}{12\pi^2} 
~ \eta^{~}_B ~ F(z) ~ (\xi^*_t)^2 \; ,
\end{equation}
where $B_B$ is the ``bag'' parameter describing the uncertainty in evaluation of the 
hadronic matrix element $\langle B^0_d|\bar{b} \gamma_{\mu} (1-\gamma_5) d
|\bar{B}^0_d\rangle$, $M_B$ and $f_B$ are the $B_d$-meson mass and decay
constant respectively, $m_t$ is the top-quark mass, $\eta^{~}_B$ denotes the
QCD correction factor, $F(z)$ stands for a slowly decreasing monotonic 
function of $z\equiv m^2_t/M^2_W$ \cite{Inami81}:
\begin{equation}
F(z) \; =\; \frac{1}{4} ~ + ~ \frac{9}{4} ~ \frac{1}{1-z} ~ - ~
\frac{3}{2} ~ \frac{1}{(1-z)^2} ~ - ~ \frac{3}{2} ~ \frac{z^2 \ln
z}{(1-z)^3} \; .
\end{equation}
In particular, $F(0)=1$, $F(1)=3/4$, and $F(\infty)=1/4$.

Next, $\Gamma^{\rm SM}_{12}$ is given as follows \cite{Hagelin81,Buras84}:
\begin{equation}
\Gamma^{\rm SM}_{12} \; =\; - ~ \frac{G^2_{\rm F} ~ B_B ~ f^2_B ~ M_B
~ m^2_b}{8\pi} ~ \left [ (\xi^*_u)^2 ~ T_u(\tilde{z}) ~ + ~ (\xi^*_c)^2
~ T_c(\tilde{z}) ~ + ~ (\xi^*_t)^2 ~ T_t(\tilde{z}) \right ] \; ,
\end{equation}
where $m_b$ is the bottom-quark mass, and $T_i(\tilde{z})$ is a function of
$\tilde{z} \equiv m^2_c/m^2_b$. Explicitly, $T_u(\tilde{z})$,
$T_c(\tilde{z})$ and $T_t(\tilde{z})$ read
\begin{eqnarray}
T_u(\tilde{z}) & = & \eta^{~}_4 ~ \tilde{z}^2 ~ \left ( 3-2 \tilde{z} \right ) ~ + ~
\frac{4}{3} ~ \eta^{~}_5 ~ \tilde{z} ~ \left (1-\tilde{z} \right )^2 \; ,
\nonumber \\
T_c(\tilde{z}) & = & \eta^{~}_4 ~ \left (1+ 2\tilde{z} \right ) \left [ \sqrt{1
-4\tilde{z}} - \left (1-\tilde{z} \right )^2 \right ] ~ - ~
\frac{4}{3} ~ \eta^{~}_5 ~ \tilde{z} ~ \left [ 2 \sqrt{1 -4\tilde{z}} -
\left (1-\tilde{z} \right )^2 \right ] \; , \nonumber \\
T_t(\tilde{z}) & = & \eta^{~}_4 ~ \left (1 + 2\tilde{z} \right ) \left (1-\tilde{z} \right
)^2 ~ - ~ \frac{4}{3} ~ \eta^{~}_5 ~ \tilde{z} ~ \left (1-\tilde{z} \right )^2
\; ,
\end{eqnarray}
in which $\eta^{~}_4$ and $\eta^{~}_5$ are two QCD correction
factors. A numerical calculation shows that $T_t (\tilde{z})$
is dominant over $T_u (\tilde{z})$ and $T_c (\tilde{z})$ in
magnitude (see below).

The $CP$ asymmetry in semileptonic $B_d$ decays (${\cal A}^{\rm
SM}_{\rm SL}$) turns out to be
\begin{equation}
{\cal A}^{\rm SM}_{\rm SL} \; = \; C_m \left [ {\rm Im} \left
(\frac{\xi_u}{\xi_t} \right )^2  T_u(\tilde{z}) ~ + ~ {\rm Im} \left
( \frac{\xi_c}{\xi_t} \right )^2  T_c(\tilde{z}) \right ] \; , 
\end{equation}
where $C_m = 3\pi m^2_b/[2 m^2_t \eta_B F(z)]$, and 
\begin{eqnarray}
{\rm Im} \left ( \frac{\xi_u}{\xi_t} \right )^2 & \approx & 
\frac{2 ~ \eta ~ \left [ \rho ~ (1 -\rho) ~ - ~ \eta^2 \right ]}{\left 
[ (1-\rho)^2 ~ + ~ \eta^2 \right ]^2} \; , \nonumber \\
{\rm Im} \left ( \frac{\xi_c}{\xi_t} \right )^2 & \approx & 
\frac{2 ~ \eta ~ (1 -\rho)}{\left 
[ (1-\rho)^2 ~ + ~ \eta^2 \right ]^2} \; .
\end{eqnarray}
The $T_t (\tilde{z})$ term, which dominates $\Gamma^{\rm SM}_{12}$,
has no contribution to the $CP$ asymmetry ${\cal A}^{\rm SM}_{\rm SL}$.

Indeed the $CP$-violating phases $\phi_1$ and $\phi_2$ can be determined from the
$CP$ asymmetries in $B_d \rightarrow J/\psi K_S$ and $B_d \rightarrow
\pi^+\pi^-$, respectively. Within the SM, 
$q/p = \xi_t/\xi^*_t$ results from the box-diagram 
calculation. From Eqs. (2.4) and (2.5), it is easy
to obtain 
\begin{eqnarray}
{\cal A}^{\rm SM}_{\psi K} & = & \sin (2\phi_1) \; \approx \; 
\frac{2 ~ \eta ~ (1-\rho)}{(1-\rho)^2 ~ + ~ \eta^2} \; , \nonumber \\
{\cal A}^{\rm SM}_{\pi\pi} & = & \sin (2\phi_2) \; \approx \;
\frac{2 ~ \eta ~ \left [ \eta^2 ~ - ~ \rho ~ (1 -\rho) \right ]}
{( \rho^2 + \eta^2) ~ \left [ (1-\rho)^2 ~ + ~ \eta^2 \right ]} \; .
\end{eqnarray}
A test of the correlation between ${\cal A}^{\rm SM}_{\rm SL}$ and
${\cal A}^{\rm SM}_{\psi K}$ or that between ${\cal A}^{\rm SM}_{\psi
K}$ and ${\cal A}^{\rm SM}_{\pi\pi}$ at $B$ factories is necessary, in order to 
find possible NP which may affect these observables in
different ways.

Let us illustrate the magnitudes of ${\cal A}^{\rm SM}_{\rm SL}$,
${\cal A}^{\rm SM}_{\psi K}$ and ${\cal A}^{\rm SM}_{\pi\pi}$ 
explicitly. The current quark masses are 
typically taken as $m_c = 1.4$ GeV, $m_b = 4.8$ GeV and $m_t = 167$
GeV; and the QCD correction factors are chosen to be $\eta^{~}_B =
0.55$, $\eta^{~}_4 = 1.15$ and $\eta^{~}_5 = 0.88$. Then one gets 
$F(z) \approx 0.55$, $T_u(\tilde{z}) \approx 0.11$, $T_c(\tilde{z})
\approx -0.11$, $T_t(\tilde{z})\approx 1.04$, and $C_m \approx
1.3\times 10^{-2}$. An analysis of current data on quark mixing and $CP$ violation
yields $\rho \approx 0.05$ and $\eta \approx 0.36$ as favored values \cite{Ali96}.
With these inputs we arrive at
\begin{equation}
{\cal A}_{\psi K}^{\rm SM} \; \approx \; 0.66 \; , ~~~~~~~~ 
{\cal A}^{\rm SM}_{\pi\pi} \; \approx \; 0.43 \; , ~~~~~~~~
{\cal A}^{\rm SM}_{\rm SL} \; \approx \; 
- 9.8 \times 10^{-4} \; .
\end{equation}
If the large errors of relevant inputs are taken into account, we find 
that the magnitude of ${\cal A}^{\rm SM}_{\rm SL}$ may change a little 
bit around $10^{-3}$, but its sign remains negative.
Clearly it is very difficult to measure such a small $CP$
asymmetry.

\section{Effects of NP on $CP$ asymmetries}
\setcounter{equation}{0}

In most extensions of the SM, NP can significantly
contribute to $M_{12}$. However, NP is not expected to significantly affect 
the direct $B$-meson decays via the tree-level $W$-mediated channels.
Thus $\Gamma_{12} = \Gamma^{\rm SM}_{12}$ holds
as a good approximation. In the presence of
NP, $M_{12}$ can be written as
\begin{equation}
M_{12} \; =\; M^{\rm SM}_{12} ~ + ~ M^{\rm NP}_{12} \; .
\end{equation}
The relative magnitude and the phase difference between $M^{\rm NP}_{12}$ and $M^{\rm 
SM}_{12}$ are unknown, while $|M_{12}| = \Delta M/2$ holds 
by definition.

It is convenient to parametrize the magnitude of $M^{\rm SM}_{12}$ as
\begin{equation}
|M^{\rm SM}_{12}| \; =\; R_{\rm SM} ~ \frac{\Delta M}{2}
\end{equation}
in subsequent discussions. The allowed range of $R_{\rm SM}$ can be 
estimated by use of Eq. (3.1) and current data; i.e.,
\begin{equation}
R_{\rm SM} \; = \; \frac{G^2_{\rm F} ~ B_B ~ f^2_B ~ M_B ~ m^2_t}{6\pi^2 ~
\Delta M} ~ \eta^{~}_B ~ F(z) ~ |\xi_t|^2 \; .
\end{equation}
Using $f_B \sqrt{B_B} = (200 \pm 40)$ MeV, $m_t = (167 \pm 6)$ GeV,  
$\Delta M = (0.464 \pm 0.018) ~ {\rm ps}^{-1}$, and 
$\eta^{~}_B = 0.55 \pm 0.01$ (see Ref. \cite{Buras96}), 
we get $R_{\rm SM} \approx (1.34 \pm 0.71) \times 10^4 ~ |\xi_t|^2$.
The large error comes primarily from the input value of $f_B
\sqrt{B_B}$, which will be improved in more delicate lattice-QCD calculations.
Since $|\xi_u|$ and $|\xi_c|$ have been measured, the most generous
constraint on $|\xi_t|$ (in the presence of NP) should be
\begin{equation}
|\xi_c| ~ - ~ |\xi_u| \; \leq \; |\xi_t| \; \leq \; |\xi_c| ~ + ~ |\xi_u| \; 
,
\end{equation}
as one can see from Fig. 1. A measurement of the rare decay
$K^+\rightarrow \pi^+ \nu \bar{\nu}$ will provide an
independent determination of (or constraint on) $|\xi_t|$. 
By use of $|\xi_u| = 0.003 \pm 0.001$ and $|\xi_c|
= 0.0087 \pm 0.0007$ \cite{PDG96,Ali96}, we get $0.1 \leq R_{\rm SM}
\leq 3.7$ as a conservative result. If only the central values of
input parameters are taken into account, then a narrower range can be obtained: 
$0.43 \leq R_{\rm SM} \leq 1.8$.

To illustrate the effect of NP on $CP$ asymmetries ${\cal A}_{\rm SL}$, 
${\cal A}_{\psi K}$ and ${\cal A}_{\pi\pi}$, we subsequently 
consider three possible NP scenarios for $M_{12}$.

\subsection{Scenario (A): ${\rm Im}(M^{\rm SM}_{12}) =0$ and ${\rm Im}(M^{\rm
NP}_{12}) \neq 0$}

In this scenario, the KM matrix is assumed to be real and
$CP$ violation arises solely from NP. Then 
$M^{\rm SM}_{12}$, $M^{\rm NP}_{12}$ and $M_{12}$ in Eq. (4.1) 
can be instructively parametrized as 
\begin{equation}
\left \{ M^{\rm SM}_{12} \; , \; M^{\rm NP}_{12} \; , \; M_{12} \right 
\} \; =\; \left \{ R_{\rm SM} \; , \; R_{\rm NP} ~ e^{{\rm i}2\theta} \; ,
\; e^{{\rm i} 2\phi} \right \} \frac{\Delta M}{2} \; ,
\end{equation}
where $R_{\rm NP}$ is a real (positive or vanishing) parameter,
$\theta$ stands for the phase of NP,
and $\phi$ is an effective phase of $B^0_d$-$\bar{B}^0_d$ mixing. In
the complex plane, $M^{\rm SM}_{12}$, $M^{\rm NP}_{12}$ and $M_{12}$
(or equivalently, $R_{\rm SM}$, $R_{\rm NP} e^{{\rm i}2\theta}$ and $e^{{\rm
i}2\phi}$) form a triangle \cite{Footnote2}, as illustrated by Fig. 2.
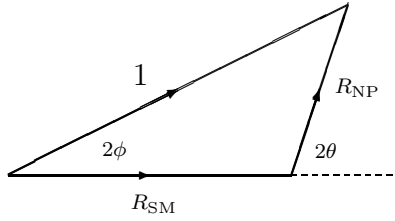
\begin{figure}[t]
\begin{picture}(400,160)(-90,210)

\put(80,300){\line(1,0){107}}
\put(80,299.5){\line(1,0){107}}
\put(80,300){\vector(1,0){53.5}}
\put(80,299.5){\vector(1,0){53.5}}
\put(135,289){\makebox(0,0){\scriptsize $R_{\rm SM}$}}
\multiput(187,300)(4,0){10}{\line(1,0){2}}

\put(187,300){\line(1,3){21.5}}
\put(187,300.5){\line(1,3){21.5}}
\put(187,299.5){\line(1,3){21.5}}
\put(187,300){\vector(1,3){10.75}}
\put(187,300.5){\vector(1,3){10.75}}
\put(187,299.5){\vector(1,3){10.75}}
\put(212,333){\makebox(0,0){\scriptsize $R_{\rm NP}$}}

\put(80,300){\line(2,1){128}}
\put(80,300.5){\line(2,1){128}}
\put(80,300){\vector(2,1){64}}
\put(80,300.5){\vector(2,1){64}}
\put(130,338){\makebox(0,0){1}}

\put(120,309){\makebox(0,0){\scriptsize $2\phi$}}
\put(200,309){\makebox(0,0){\scriptsize $2\theta$}}
\end{picture}
\vspace{-2.5cm}
\caption{\small Triangular relation of $M^{\rm SM}_{12}$, $M^{\rm
NP}_{12}$ and $M_{12}$ (rescaled by $\Delta M/2$) in scenario (A).}
\end{figure}
By use of the triangular relation, $R_{\rm NP}$ can be expressed
as
\begin{equation}
R_{\rm NP} \; =\; - R_{\rm SM} ~ \cos (2\theta) ~
\pm ~ \sqrt{1 - R^2_{\rm SM} ~ \sin^2(2\theta)} \; .
\end{equation}
We find that two solutions exist for $R_{\rm NP}$, 
corresponding to ($\pm$) signs on the right-hand 
side of Eq. (4.6). Since the magnitude
of $R_{\rm SM}$ has been constrained to some extent, we are able to obtain
the allowed $(\theta, R_{\rm NP})$ parameter space numerically.

The $CP$-violating phase $\phi$ can be measured from the $CP$
asymmetry in $B_d\rightarrow J/\psi K_S$. With the help of Eq. (2.4),
we obtain
\begin{equation}
{\cal A}_{\psi K} \; =\; \sin (2\phi) \; =\; R_{\rm NP} ~ \sin(2\theta) \; .
\end{equation}
Of course, $|{\cal A}_{\psi K}| \leq 1$ holds for $R_{\rm NP}$ and
$\theta$ to take values allowed by Eq. (4.6).

The $CP$ asymmetry in $B_d \rightarrow \pi^+\pi^-$ is 
simply related to ${\cal A}_{\psi K}$. By use of Eq. (2.5), we find
\begin{equation}
{\cal A}_{\pi\pi} \; = - \sin (2\phi) \; = \; - ~ {\cal A}_{\psi K} \; .
\end{equation}
Such a linear correlation between ${\cal A}_{\psi K}$ and ${\cal A}_{\pi\pi}$
is a straightforward consequence of the superweak scenario of NP
considered here.

The $CP$ asymmetry in semileptonic $B_d$ decays, defined in Eq. (2.3), is given by
\begin{equation}
{\cal A}_{\rm SL} \; = \; C_m ~ R_{\rm SM} ~ R_{\rm NP} ~ \sin (2\theta)
\left [ \left ( \frac{\tilde{\xi}_u}{\tilde{\xi}_t} \right )^2  T_u(\tilde{z}) ~ + ~ 
\left ( \frac{\tilde{\xi}_c}{\tilde{\xi}_t} \right )^2  T_c(\tilde{z}) ~ + ~
T_t(\tilde{z}) \right ] \; ,
\end{equation}
where $\tilde{\xi}_u$, $\tilde{\xi}_c$ and $\tilde{\xi}_t$
denote the real KM factors in scenario (A).
Different from the SM 
result in Eq. (3.5), whose magnitude is associated only with $T_u
(\tilde{z}) \sim O(0.1)$ and $T_c (\tilde{z}) \sim O(0.1)$, 
here the magnitude of ${\cal A}_{\rm SL}$ is dominated by $T_t
(\tilde{z}) \sim O(1)$. Hence $|{\cal A}_{\rm SL} / {\cal A}^{\rm SM}_{\rm SL}|
\geq 10$ is possible within an appropriate $(\theta, R_{\rm NP})$ parameter space.
We shall present numerical estimates for ${\cal A}_{\rm SL}$ and
${\cal A}_{\psi K}$ in the next section.

\subsection{Scenario (B): ${\rm Im}(M^{\rm NP}_{12}) =0$ and ${\rm Im}(M^{\rm
SM}_{12}) \neq 0$}

In this scenario the NP contribution is of the so-called
``real superweak'' type \cite{Wolfenstein93,Wolfenstein87}. The phase of $M^{\rm
SM}_{12}$ comes from the KM matrix
and amounts to $2\phi_1$, as one can see from Eq. (3.1) and Fig. 1.
For simplicity, we parametrize $M^{\rm SM}_{12}$, 
$M^{\rm NP}_{12}$ and $M_{12}$ as follows:
\begin{equation}
\left \{ M^{\rm SM}_{12} \; , \; M^{\rm NP}_{12} \; , \; M_{12} \right 
\} \; =\; \left \{ R_{\rm SM} ~ e^{{\rm i} 2\phi_1} \; , \;
R_{\rm NP} \; , \; e^{{\rm i} 2\phi} \right \} \frac{\Delta M}{2} \; ,
\end{equation}
where $R_{\rm NP}$ is a real parameter, and $\phi$ denotes the effective
phase of $B^0_d$-$\bar{B}^0_d$ mixing. Clearly 
$M^{\rm SM}_{12}$, $M^{\rm NP}_{12}$ and $M_{12}$ (or equivalently,
$R_{\rm SM} e^{{\rm i} 2\phi_1}$, $R_{\rm NP}$ and $e^{{\rm i} 2\phi}$)
form a triangle in the complex plane, as illustrated by Fig. 3.
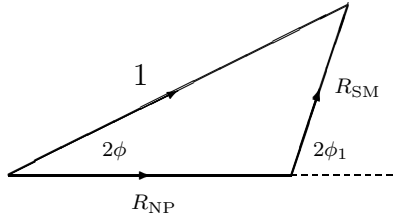
\begin{figure}[t]
\begin{picture}(400,160)(-90,210)

\put(80,300){\line(1,0){107}}
\put(80,299.5){\line(1,0){107}}
\put(80,300){\vector(1,0){53.5}}
\put(80,299.5){\vector(1,0){53.5}}
\put(135,289){\makebox(0,0){\scriptsize $R_{\rm NP}$}}
\multiput(187,300)(4,0){10}{\line(1,0){2}}

\put(187,300){\line(1,3){21.5}}
\put(187,300.5){\line(1,3){21.5}}
\put(187,299.5){\line(1,3){21.5}}
\put(187,300){\vector(1,3){10.75}}
\put(187,300.5){\vector(1,3){10.75}}
\put(187,299.5){\vector(1,3){10.75}}
\put(212,333){\makebox(0,0){\scriptsize $R_{\rm SM}$}}

\put(80,300){\line(2,1){128}}
\put(80,300.5){\line(2,1){128}}
\put(80,300){\vector(2,1){64}}
\put(80,300){\vector(2,1){64}}
\put(130,338){\makebox(0,0){1}}

\put(120,309){\makebox(0,0){\scriptsize $2\phi$}}
\put(202,309){\makebox(0,0){\scriptsize $2\phi_1$}}
\end{picture}
\vspace{-2.5cm}
\caption{\small Triangular relation of $M^{\rm SM}_{12}$, $M^{\rm
NP}_{12}$ and $M_{12}$ (rescaled by $\Delta M/2$) in scenario (B).}
\end{figure}
In terms of $R_{\rm SM}$ and $\phi_1$, $R_{\rm NP}$ can be written
as \cite{Wolfenstein93}
\begin{equation}
R_{\rm NP} \; =\; - R_{\rm SM} ~ \cos (2\phi_1) ~
\pm ~ \sqrt{1 - R^2_{\rm SM} ~ \sin^2(2\phi_1)} \; . 
\end{equation}
We see that there are two solutions for $R_{\rm NP}$,
corresponding to $(\pm)$ signs on the right-hand side of
Eq. (4.11).

The $CP$-violating phase $\phi$ can be determined from the $CP$
asymmetry in $B_d\rightarrow J/\psi K_S$:
\begin{equation}
{\cal A}_{\psi K} \; =\; \sin (2\phi) \; =\; R_{\rm SM} ~ \sin
(2\phi_1) \; .
\end{equation}
Of course, $|{\cal A}_{\psi K}| \leq 1$ 
holds if $R_{\rm SM}$ and $\phi_1$ vary in their allowed regions.

The $CP$ asymmetry in $B_d \rightarrow \pi^+\pi^-$ reads:
\begin{equation}
{\cal A}_{\pi\pi} \; = \; - \sin 2 (\phi + \phi_3) \; = \; 
R_{\rm SM} ~ \sin (2\phi_2) ~ - ~ R_{\rm NP} ~ \sin (2\phi_3) \; ,
\end{equation}
where $\phi_2$ and $\phi_3$ are the second and third angles of the
unitarity triangle (see Fig. 1). The correlation between ${\cal
A}_{\psi K}$ and ${\cal A}_{\pi\pi}$ does exist, because of $\phi_1 +
\phi_2 + \phi_3 = \pi$. 
In the absence of NP (i.e., $R_{\rm NP} = 0$ and $R_{\rm SM} = 1$), Eqs. (4.12) and (4.13) 
will be simplified to the SM results as given by Eq. (3.7).

The $CP$ asymmetry ${\cal A}_{\rm SL}$ in scenario (B) turns out to be
\small
\begin{eqnarray}
{\cal A}_{\rm SL} & = & C_m ~ R_{\rm SM} ~ R_{\rm NP} ~ \left [ {\rm Im} 
\left ( \frac{\xi_u}{|\xi_t|}\right 
)^2 T_u (\tilde{z}) ~ + ~ {\rm Im} \left ( \frac{\xi_c}{|\xi_t|}\right )^2 T_c
(\tilde{z}) ~ + ~ {\rm Im} \left ( \frac{\xi_t}{|\xi_t|}\right )^2
T_t(\tilde{z}) \right ] \nonumber \\
&  & + ~ C_m ~ R^2_{\rm SM} ~ \left [ {\rm Im} \left
( \frac{\xi_u}{\xi_t} \right )^2 T_u (\tilde{z}) ~ + ~ {\rm Im} \left
( \frac{\xi_c}{\xi_t} \right )^2 T_c (\tilde{z}) \right ] \; .
\end{eqnarray}
\normalsize
One can see that ${\cal A}_{\rm SL}$ consists of two terms:
the first comes from the interference between $M^{\rm
SM}_{12}$ and $M^{\rm NP}_{12}$ in $M_{12}$, while the second is
purely a SM-like contribution from $M^{\rm SM}_{12}$ itself
(see Eq. (3.5) for comparison). Within a suitable parameter space,
the magnitude of ${\cal A}_{\rm SL}$ should be dominated 
by the term associated with $T_t (\tilde{z})$; thus
$|{\cal A}_{\rm SL} / {\cal A}^{\rm SM}_{\rm SL}| \geq 10$ is possible
in this NP scenario.

\subsection{Scenario (C): ${\rm Im}(M^{\rm NP}_{12}) \neq 0$ and ${\rm Im}(M^{\rm
SM}_{12}) \neq 0$}

This is a quite general NP scenario which can accommodate both
scenarios (A) and (B). As done before, we parametrize $M^{\rm SM}_{12}$, 
$M^{\rm NP}_{12}$ and $M_{12}$ in the following way:
\begin{equation}
\left \{ M^{\rm SM}_{12} \; , \; M^{\rm NP}_{12} \; , \; M_{12} \right 
\} \; =\; \left \{ R_{\rm SM} ~ e^{{\rm i} 2\phi_1} \; , \;
R_{\rm NP} ~ e^{{\rm i} 2\theta} \; , \; e^{{\rm i} 2\phi} \right \}
\frac{\Delta M}{2} \; ,
\end{equation}
where $R_{\rm NP}$ is a real (positive or vanishing) parameter, 
$\theta$ represents the NP phase, 
and $\phi$ denotes the effective
phase of $B^0_d$-$\bar{B}^0_d$ mixing. In this case,
$M^{\rm SM}_{12}$, $M^{\rm NP}_{12}$ and $M_{12}$ (or equivalently,
$R_{\rm SM} e^{{\rm i} 2\phi_1}$, $R_{\rm NP} e^{{\rm i} 2\theta}$ and $e^{{\rm i} 2\phi}$)
form a triangle in the complex plane, as illustrated by Fig. 4.
\begin{figure}[t]
\begin{picture}(400,160)(-90,210)

\multiput(60,300)(4,0){40}{\line(1,0){2}}

\put(80,300){\line(1,3){22}}
\put(80,300.5){\line(1,3){22}}
\put(80,299.5){\line(1,3){22}}
\put(80,300){\vector(1,3){11}}
\put(80,300.5){\vector(1,3){11}}
\put(80,299.5){\vector(1,3){11}}
\put(81,336){\makebox(0,0){1}}

\put(102,366){\line(3,-2){47}}
\put(102,366.5){\line(3,-2){47}}
\put(149,334.5){\vector(-3,2){24}}
\put(149,335){\vector(-3,2){24}}
\put(140,355){\makebox(0,0){\scriptsize $R_{\rm NP}$}}

\multiput(149,334.5)(3,-2){18}{\circle*{0.3}}

\put(80,300){\line(2,1){69}}
\put(80,300.5){\line(2,1){69}}
\put(80,300){\vector(2,1){35}}
\put(80,300){\vector(2,1){35}}
\put(113,330){\makebox(0,0){\scriptsize $R_{\rm SM}$}}

\put(119,308){\makebox(0,0){\scriptsize $2\phi_1$}}
\put(203,308){\makebox(0,0){\scriptsize $2\theta$}}
\end{picture}
\vspace{-2.6cm}
\caption{\small Triangular relation of $M^{\rm SM}_{12}$, $M^{\rm
NP}_{12}$ and $M_{12}$ (rescaled by $\Delta M/2$) in scenario (C).}
\end{figure}
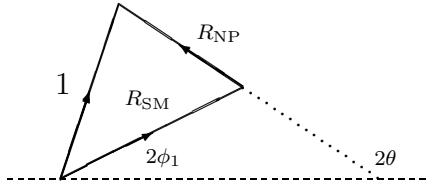
In terms of $R_{\rm SM}$, $\phi_1$ and $\theta$, $R_{\rm NP}$ can be expressed
as
\begin{equation}
R_{\rm NP} \; =\; - R_{\rm SM} ~ \cos 2 (\theta - \phi_1) ~
\pm ~ \sqrt{1 - R^2_{\rm SM} ~ \sin^2 2 (\theta - \phi_1)} \; . 
\end{equation}
We observe that $R_{\rm NP}$ depends on the phase difference between
$\theta$ and  $\phi_1$. Also there exist two solutions for $R_{\rm NP}$,
corresponding to $(\pm)$ signs on the right-hand side of
Eq. (4.16). In the special case $\theta = \phi_1$, one arrives at $R_{\rm NP} =
-R_{\rm SM} \pm 1$.

It is straightforward to derive the $CP$ asymmetry ${\cal A}_{\psi K}$
in $B^0_d$ vs $\bar{B}^0_d \rightarrow J/\psi K_S$ modes:
\begin{equation}
{\cal A}_{\psi K} \; =\; \sin (2\phi) \; =\; R_{\rm SM} ~ \sin
(2\phi_1) ~ + ~ R_{\rm NP} ~ \sin (2\theta)  \; .
\end{equation}
Since $R_{\rm NP}$, $R_{\rm SM}$ and $\phi_1$, $\theta$ are dependent
on one another through Eq. (4.16), $|{\cal A}_{\psi K}| \leq 1$
is always guaranteed within the allowed parameter space.

In scenario (C), the $CP$ asymmetry in $B_d \rightarrow \pi^+\pi^-$ is
given as
\begin{equation}
{\cal A}_{\pi\pi} \; = \; - \sin 2 (\phi + \phi_3) \; = \; 
R_{\rm SM} ~ \sin (2\phi_2) ~ - ~ R_{\rm NP} ~ \sin 2 ( \theta + \phi_3) \; ,
\end{equation}
where $\phi_2$ and $\phi_3$ are two inner angles of the
unitarity triangle in Fig. 1. In comparison with scenarios (A) and
(B), here the correlation between ${\cal
A}_{\psi K}$ and ${\cal A}_{\pi\pi}$ becomes more complicated.
We see that the results in Eqs. (4.12) and (4.13) can be
respectively reproduced from Eqs. (4.17) and (4.18) if $\theta =0$ is taken.

The $CP$ asymmetry ${\cal A}_{\rm SL}$ in scenario (C) reads
\small
\begin{eqnarray}
{\cal A}_{\rm SL} & = & C_m ~ R_{\rm SM} ~ R_{\rm NP} ~ \left [ {\rm Im} 
\left ( \frac{\xi_u}{|\xi_t|}\right 
)^2 T_u (\tilde{z}) ~ + ~ {\rm Im} \left ( \frac{\xi_c}{|\xi_t|}\right )^2 T_c
(\tilde{z}) ~ + ~ {\rm Im} \left ( \frac{\xi_t}{|\xi_t|}\right )^2
T_t(\tilde{z}) \right ] \cos (2\theta) \nonumber \\
&  & + ~ C_m ~ R^2_{\rm SM} ~ \left [ {\rm Im} \left
( \frac{\xi_u}{\xi_t} \right )^2 T_u (\tilde{z}) ~ + ~ {\rm Im} \left
( \frac{\xi_c}{\xi_t} \right )^2 T_c (\tilde{z}) \right ] \; + \; \\ 
&  & C_m ~ R_{\rm SM} ~ R_{\rm NP} ~ \left [ {\rm Re} 
\left ( \frac{\xi_u}{|\xi_t|}\right 
)^2 T_u (\tilde{z}) ~ + ~ {\rm Re} \left ( \frac{\xi_c}{|\xi_t|}\right )^2 T_c
(\tilde{z}) ~ + ~ {\rm Re} \left ( \frac{\xi_t}{|\xi_t|}\right )^2
T_t(\tilde{z}) \right ] \sin (2\theta) \; . \nonumber 
\end{eqnarray}
\normalsize
Clearly the second term of ${\cal A}_{\rm SL}$ comes purely from
$M^{\rm SM}_{12}$ itself and its magnitude is expected to be of $O(10^{-3})$.
The first and third terms of ${\cal A}_{\rm SL}$ arise from the
interference between $M^{\rm SM}_{12}$ and $M^{\rm NP}_{12}$;
but they depend on nonvanishing ${\rm Im} (M^{\rm SM}_{12})$ and
${\rm Im} (M^{\rm NP}_{12})$, respectively. Thus ${\cal A}_{\rm SL}
(\theta =0)$ is just the result given by Eq. (4.14) for scenario (B).
For appropriate values of $\theta$ and $\phi_1$,
magnitudes of both the first and third terms of ${\cal A}_{\rm SL}$ may
be at the percent level. To obtain $|{\cal A}_{\rm SL}| \sim O(10^{-2})$,
however, there should not be large cancellation 
between two dominant terms in Eq. (4.19).

\section{Numerical estimates of $CP$ asymmetries}
\setcounter{equation}{0}

For the purpose of illustration,
let us estimate the magnitudes of $CP$ asymmetries obtained from the
above NP scenarios. Since scenario (C) involves several unknown
parameters (though they are related to one another through
Eq. (4.16)), a numerical analysis of its allowed parameter space would 
be complicated and less instructive \cite{Goto96}. 
Hence we shall only concentrate on scenarios (A) and (B) in the
following.

\subsection{Results for scenario (A)}

Following the spirit of the Wolfenstein parametrization \cite{Wolfenstein83},
the real KM matrix $\tilde{V}$ in scenario (A) can be parametrized in terms of
three independent parameters $\tilde{\lambda}$, $\tilde{A}$ and
$\tilde{\rho}$. Taking $\tilde{V}_{us} =\tilde{\lambda}$, $\tilde{V}_{cb} =
\tilde{A} \tilde{\lambda}^2$, $\tilde{V}_{ub} = \tilde{A}
\tilde{\lambda}^3 \tilde{\rho}$ and using the orthogonality
conditions of $\tilde{V}$, one can derive the other six matrix
elements. In particular, we get $\tilde{V}_{td} \approx \tilde{A}
\tilde{\lambda}^3 (1- \tilde{\rho})$. In view of current
data on $|\tilde{V}_{us}|$, $|\tilde{V}_{cb}|$ and
$|\tilde{V}_{ub}/\tilde{V}_{cb}|$ (see Ref. \cite{PDG96}), we find $\tilde{\lambda} 
\approx 0.22$, $\tilde{A} \approx 0.8$ and $\tilde{\rho} \approx \pm
0.36$ typically. The sign 
ambiguity of $\tilde{\rho}$ may affect the allowed parameter space of 
NP as well as the $CP$ asymmetries of $B$-meson decays, as one can see 
later on.

The size of $R_{\rm SM}$ depends on the real KM factor
$\tilde{\xi}_t$. With the help of Eq. (4.3), we get
$R_{\rm SM} \approx 1.8$ for $\tilde{\rho} \approx -0.36$ and $R_{\rm SM} \approx
0.43$ for $\tilde{\rho} \approx +0.36$. The corresponding
results of $R_{\rm NP}$, changing with $\theta$, can be obtained from
Eq. (4.6). We find that for $\tilde{\rho} \approx +0.36$ the $(-)$
solution of $R_{\rm NP}$ is not allowed. 
The allowed $(\theta, R_{\rm NP})$ parameter space is shown in Fig. 5.
\begin{figure}
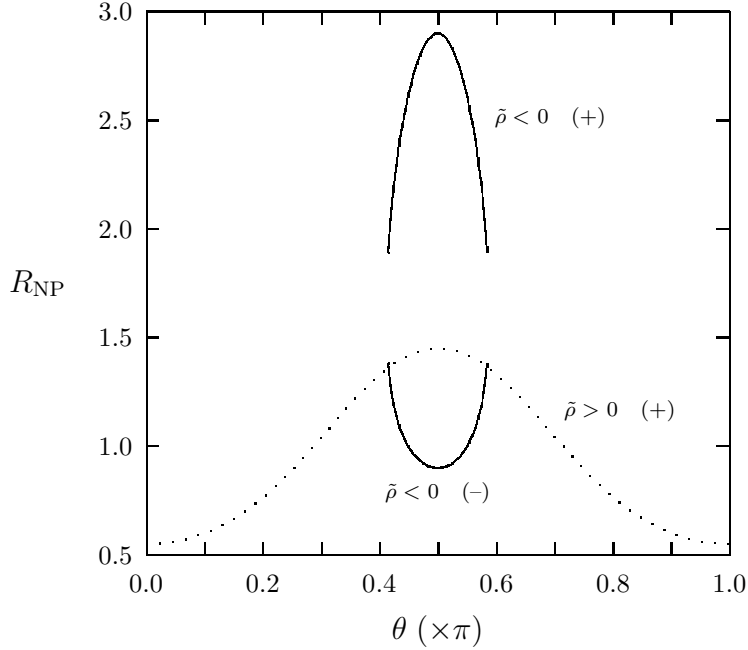

\setlength{\unitlength}{0.240900pt}
\ifx\plotpoint\undefined\newsavebox{\plotpoint}\fi
\sbox{\plotpoint}{\rule[-0.200pt]{0.400pt}{0.400pt}}%

\vspace{0.4cm}
\caption{\small Illustrative plot for the $(\theta, R_{\rm NP})$ parameter
space in scenario (A).}
\end{figure}
\begin{figure}
\setlength{\unitlength}{0.240900pt}
\ifx\plotpoint\undefined\newsavebox{\plotpoint}\fi
\sbox{\plotpoint}{\rule[-0.200pt]{0.400pt}{0.400pt}}%
%
\vspace{0.4cm}
\caption{\small Illustrative plot for the $CP$ asymmetries 
${\cal A}_{\rm SL}$ and ${\cal A}_{\psi K}$ in scenario (A).}
\end{figure}

The $CP$ asymmetries ${\cal A}_{\rm SL}$ and ${\cal A}_{\psi K}$ (or
${\cal A}_{\pi\pi}$) in this scenario can then be calculated by use of the
above obtained parameter space. For simplicity, we express the KM
factors $\tilde{\xi}_u$, $\tilde{\xi}_c$ and $\tilde{\xi}_t$ in
Eq. (4.9) in terms of $\tilde{\lambda}$, $\tilde{A}$ and
$\tilde{\rho}$. Then the correlation between ${\cal A}_{\rm SL}$ and 
${\cal A}_{\psi K}$ reads
\begin{equation}
{\cal A}_{\rm SL} \; \approx \; C_m ~ R_{\rm SM} 
\left [ \frac{\tilde{\rho}^2}{(1 - \tilde{\rho})^2} ~ T_u(\tilde{z}) ~ + ~ 
\frac{1}{(1 - \tilde{\rho})^2} ~ T_c(\tilde{z}) ~ + ~
T_t(\tilde{z}) \right ] {\cal A}_{\psi K} \; 
\end{equation}
with $\tilde{\rho} \approx \pm 0.36$. The results of ${\cal A}_{\rm
SL}$ and ${\cal A}_{\psi K}$, changing with the $CP$-violating phase
$\theta$, are depicted in Fig. 6. Two remarks are in order.

a) With the typical inputs mentioned above, we have found ${\cal
A}_{\rm SL}/ {\cal A}_{\psi K} \approx 0.023$ for $\tilde{\rho}
\approx -0.36$ and ${\cal A}_{\rm SL} /{\cal A}_{\psi K} \approx
0.0042$ for $\tilde{\rho} \approx +0.36$. Because of $|{\cal A}_{\psi
K}| \leq 1$, only the former case is likely to lead the magnitude of 
${\cal A}_{\rm SL}$ to the percent level.

b) In both cases, however, the $CP$ asymmetry ${\cal A}_{\psi K}$ may
take promising values (e.g., $|{\cal A}_{\psi K}| \geq 0.5$). The 
correlation between ${\cal A}_{\psi K}$ and ${\cal A}_{\pi\pi}$, i.e.,
${\cal A}_{\pi\pi} = - {\cal A}_{\psi K}$, is particularly interesting
in this superweak scenario of NP. If this correlation and the one
between ${\cal A}_{\rm SL}$ and ${\cal A}_{\psi K}$ can be measured
at a $B$-meson factory, they will provide a strong constraint on the 
underlying NP in $B^0_d$-$\bar{B}^0_d$ mixing.

\subsection{Results for scenario (B)}

To calculate $R_{\rm NP}$ with the help of Eq. (4.11), we should first 
estimate $R_{\rm SM}$ by use of Eq. (4.3). 
The magnitude of $R_{\rm SM}$ depends on $|\xi_t|$:
\begin{equation}
|\xi_t| \; \approx \; A \lambda^3 \sqrt{(1-\rho)^2 + \eta^2} \; \approx \;
A \lambda^3 \sqrt{1 - 2\rho + \chi^2} \; ,
\end{equation}
where $\chi$ has been defined in Eq. (2.7). From current data on
$|V_{us}|$, $|V_{cb}|$ and $|V_{ub}/V_{cb}|$, we get 
$\lambda \approx 0.22$, $A \approx 0.8$ and $\chi \approx 0.36$.
The most generous range of 
$\rho$, due to the presence of NP, should be $-\chi \leq \rho \leq +
\chi$. We find that the resultant region of $|\xi_t|$ is just the one
given in Eq. (4.4). 
Note that $\phi_1$ is also a function of $\rho$ and $\chi$, i.e.,
\begin{equation}
\tan \phi_1 \; \approx \; \pm ~ \frac{\sqrt{\chi^2 - \rho^2}}{1-\rho} \; ,
\end{equation}
thus it cannot
take arbitrary values from $0$ to $\pi$. For this reason, it is more
convenient to calculate the $(\rho, R_{\rm NP})$ parameter space of
scenario (B) by taking $\rho \in [-0.36, +0.36]$. 
We get $0^{\circ}\leq \phi_1 \leq 20.1^{\circ}$
or $158.9^{\circ} \leq \phi_1 \leq 180^{\circ}$, corresponding to the
$(+)$ or $(-)$ sign of $\tan\phi_1$. The allowed region of 
$R_{\rm NP}$ changing with $\rho$ is shown in Fig. 7.
\begin{figure}
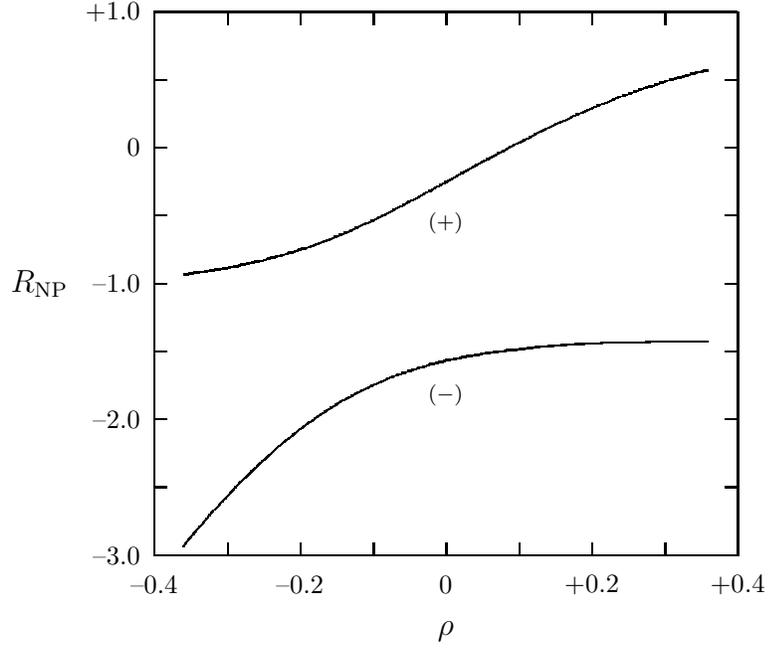

\setlength{\unitlength}{0.240900pt}
\ifx\plotpoint\undefined\newsavebox{\plotpoint}\fi
\sbox{\plotpoint}{\rule[-0.200pt]{0.400pt}{0.400pt}}%

\vspace{0.4cm}
\caption{\small Illustrative plot for the $(\rho, R_{\rm NP})$ parameter
space in scenario (B).}
\end{figure}
\begin{figure}
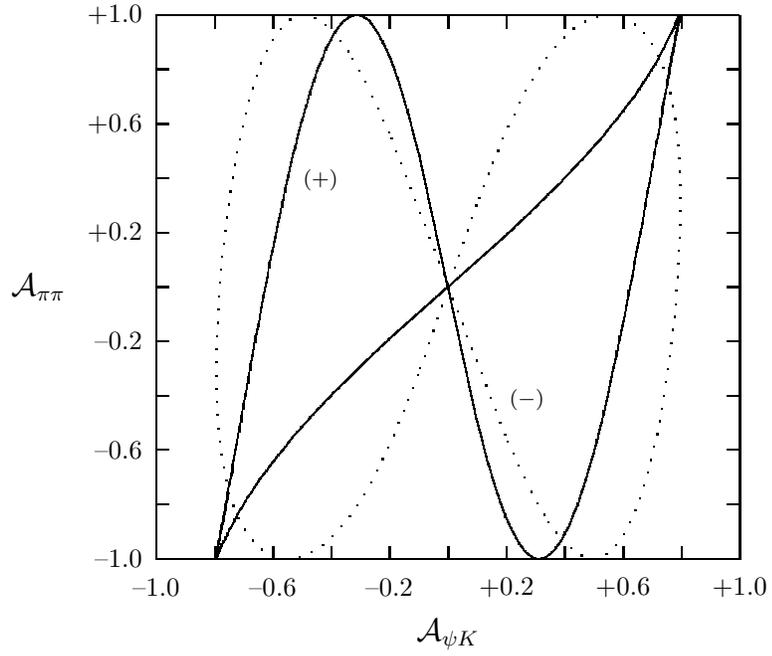

\setlength{\unitlength}{0.240900pt}
\ifx\plotpoint\undefined\newsavebox{\plotpoint}\fi
\sbox{\plotpoint}{\rule[-0.200pt]{0.400pt}{0.400pt}}%
%
\vspace{0.4cm}
\caption{\small Illustrative plot for the correlation between ${\cal
A}_{\psi K}$ and ${\cal A}_{\pi\pi}$ in scenario (B).}
\end{figure}
\begin{figure}
\setlength{\unitlength}{0.240900pt}
\ifx\plotpoint\undefined\newsavebox{\plotpoint}\fi
\sbox{\plotpoint}{\rule[-0.200pt]{0.400pt}{0.400pt}}%
%
\vspace{0.4cm}
\caption{\small Illustrative plot for the $CP$ asymmetries ${\cal A}_{\rm
SL}$ and ${\cal A}_{\psi K}$ in scenario (B).}
\end{figure}

Now we calculate the $CP$ asymmetries ${\cal A}_{\psi K}$, ${\cal
A}_{\pi\pi}$ and ${\cal A}_{\rm SL}$ by use of Eqs. (4.12), (4.13) and 
(4.14), respectively. In terms of the parameters $\rho$ and $\chi$,
the correlation between ${\cal A}_{\rm \pi\pi}$ and ${\cal A}_{\psi K}$
can be given as
\begin{equation}
{\cal A}_{\pi\pi} \; \approx \; - \left [ \frac{\rho -\chi^2}{\chi^2 (1-\rho)} ~ + ~
\frac{\rho ~ (1-2\rho + \chi^2)}{\chi^2 (1-\rho)} ~ \frac{R_{\rm
NP}}{R_{\rm SM}} \right ] {\cal A}_{\psi K} \; .
\end{equation}
Similarly we obtain the correlation between ${\cal A}_{\rm SL}$ and
${\cal A}_{\psi K}$ as follows:
\begin{equation}
{\cal A}_{\rm SL} \; \approx \; C_m ~ f(\rho, \chi) ~ {\cal A}_{\psi
K} \; ,
\end{equation}
where 
\begin{eqnarray}
f(\rho, \chi) & \equiv & R_{\rm SM} \left [ \frac{\rho - \chi^2}{(1-\rho) ~ (1-2\rho +
\chi^2)} ~ T_u (\tilde{z}) ~ + ~ \frac{1}{1-2\rho + \chi^2} ~ T_c
(\tilde{z}) \right ] ~ + ~ \nonumber \\
&  & R_{\rm NP} \left [ \frac{\rho}{1 -\rho} 
~ T_u (\tilde{z}) ~ - ~ T_t (\tilde{z}) \right ] \; .
\end{eqnarray}
Note that the term associated with the KM factor ${\rm Im} (\xi_c
/|\xi_t|)^2$ in Eq. (4.14) does not appear in $f(\rho, \chi)$, because this
factor approximately vanishes in the Wolfenstein parametrization.
The numerical results for three $CP$ asymmetries are shown in Figs. 8
and 9, where relevant inputs have been used to get the $(\rho, R_{\rm
NP})$ parameter space in Fig. 7. Three remarks are in order.

a) The two solutions of $R_{\rm NP}$ lead to identical results for the 
$CP$ asymmetry ${\cal A}_{\psi K}$. The reason is simply that ${\cal
A}_{\psi K}$ depends only on $R_{\rm SM}$ and $\phi_1$, as given in
Eq. (4.12).

b) The correlation between ${\cal A}_{\pi\pi}$ and ${\cal A}_{\psi K}$ 
is complicated here, compared with that in scenario (A) where the superweak
relation ${\cal A}_{\pi\pi} = - {\cal A}_{\psi K}$ holds. We
observe that both $CP$ asymmetries can take the same sign and
promising magnitudes in scenario (B) \cite{Footnote3}.

c) Due to their correlation, behaviors of ${\cal A}_{\rm SL}$ and ${\cal
A}_{\psi K}$ changing with $\rho$ are similar.
The magnitude of ${\cal A}_{\rm SL}$ can reach the percent level 
for appropriate values of $\rho$, if $R_{\rm NP}$ takes its $(-)$ solution.

\section{Concluding remarks}
\setcounter{equation}{0}

Importance of measuring the $CP$ asymmetry
(${\cal A}_{\rm SL}$) in semileptonic
$B_d$ decays, which can serve as a sensitive probe of NP in
$B^0_d$-$\bar{B}^0_d$ mixing, has been highly stressed. 
We have taken three NP scenarios into account, and proposed a simple 
framework for analyzing their effects on 
${\cal A}_{\rm SL}$ and other $CP$ asymmetries.
Some numerical estimates have also been made to illustrate possible
enhancement of $CP$ asymmetries in the presence of NP. 
We find that the magnitude of ${\cal A}_{\rm SL}$ at
the percent level cannot be excluded within the appropriate parameter space. The
correlation of ${\cal A}_{\rm SL}$ with the $CP$ asymmetries in $B_d
\rightarrow J/\psi K_S$ and $B_d \rightarrow \pi^+\pi^-$ (i.e., ${\cal 
A}_{\psi K}$ and ${\cal A}_{\pi\pi}$) have 
been discussed. Measuring the correlation between ${\cal
A}_{\psi K}$ and ${\cal A}_{\pi\pi}$ may partly testify or abandon 
the superweak model of $CP$ violation which was proposed long time ago
\cite{Wolfenstein64}, at least within the $B^0_d$-$\bar{B}^0_d$ system 
\cite{Xing97}.

If the $CP$ asymmetry ${\cal A}_{\rm SL}$ is really of the
order $10^{-2}$, it should be detected at the forthcoming $B$-meson
factories, where as many as $10^8$ $B^0_d\bar{B}^0_d$ events will be
produced at the $\Upsilon (4S)$ resonance in about one year. Indeed the experimental
sensitivities to ${\cal A}_{\rm SL}$ are expected to be
well within few percent for either single lepton or dilepton asymmetry
measurements \cite{Yamamoto97}, if the number of $B^0_d\bar{B}^0_d$ events is larger
than $10^7$ or so.

\vspace{0.5cm}
\begin{flushleft}
{\Large\bf Acknowledgments}
\end{flushleft}

This work was supported in part by the Grant-in-Aid for Scientific
Research on Priority Areas ({\it Physics of $CP$ Violation}) from the
Ministry of Education, Science and Culture of Japan.
Also AIS likes to thank the Daiko Foundation for a
partial support to his research, and ZZX is indebted to the Japan Society
for the Promotion of Science for its financial support.


\begin{thebibliography}{99}
\bibitem{Footnote1} This scenario is essentially different from the one
proposed by Cocolicchio and Maiani \cite{Maiani92}. In the latter, the 
rate of $B^0_d$-$\bar{B}^0_d$ mixing remains to be governed completely by the
SM with a real KM matrix. Hence it seems difficult to
simultaneously accommodate current data on $|V_{ub}/V_{cb}|$ and
$B^0_d$-$\bar{B}^0_d$ mixing.
\bibitem{Footnote2} Here we follow the procedure of Soares and Wolfenstein
\cite{Wolfenstein93} to express Eq. (4.1) geometrically, i.e., in the
form of a triangle in the complex plane.
\bibitem{Footnote3} Based on the same NP scenario as our scenario
(B), Soares and Wolfenstein have studied  the
correlation between the $CP$ asymmetries in $B_d \rightarrow J/\psi
K_S$ and $B_d \rightarrow \pi^+\pi^-$ \cite{Wolfenstein93}. 
The difference between their and our
numerical results can be traced back to different values of input
parameters.
\end{thebibliography}
\end{document}